\documentclass[11pt]{article}
\usepackage{amsmath,amsfonts,amssymb,latexsym}

\newtheorem{theorem}{Theorem}[section]

\numberwithin{equation}{section}
\def\be{\begin{equation}}
\def\ee{\end{equation}}
\def\bq{\begin{eqnarray}}
\def\eq{\end{eqnarray}}
\def\beq{\begin{eqnarray*}}
\def\eeq{\end{eqnarray*}}

\def\r{\rho}
\def\a{\alpha}
\def\b{\beta}
\def\g{\gamma}
\def\G{\Gamma}
\def\d{\delta}

\def\m{\mu}

\def\pa{\partial}
\def\z{\zeta}
\def\e{\eta}

\def\ep{\epsilon}
\newcommand{\GA}{\alpha}
\newcommand{\GB}{\beta}
\newcommand{\GG}{\gamma}
\newcommand{\GD}{\delta}
\newcommand{\GE}{\epsilon}

\newcommand{\GZ}{\zeta}

\begin{document}
\title{\huge{\textsc{The regular state in higher-order gravity}}}
\author{{\Large\textsc{Spiros Cotsakis\footnote{On leave from the University of the Aegean, 83200 Samos, Greece.}\,\,$^{1}$\thanks{\texttt{skot@aegean.gr}}, 
Seifedine Kadry$^{1}$\thanks{\texttt{Seifedine.Kadry@aum.edu.kw}},}}\\ {\Large\textsc{Dimitrios Trachilis$^{2}$\thanks{\texttt{dtrachilis@aegean.gr}}}} \\
$^{1}$Department of Mathematics, \\ American University of the Middle East\\P. O. Box 220 Dasman, 15453, Kuwait \\
$^{2}$Research group of Geometry, Dynamical Systems and Cosmology,\\
University of the Aegean, Karlovassi 83200, Samos, Greece.}
\maketitle
\begin{abstract}
\noindent We consider the higher-order gravity theory derived from the quadratic lagrangian $R+\ep R^2$ in vacuum as a first-order (ADM-type) system with constraints, and build time developments of  solutions of an initial value formulation of the theory. We show that all such solutions, if analytic, contain the right number of free functions to qualify as  general solutions of the theory. We further show that any regular analytic solution which satisfies the constraints and the evolution equations can be  given in the form of an asymptotic formal power series expansion.
\end{abstract}

\section{Introduction}
\noindent Ever since the fundamental realization that Einstein's equations can be equivalently studied as a geometric, nonlinear system of evolution equations with constraints, there has been a considerable stream of problems in general relativity  of a classical dynamical nature crucially depending on the notion of evolution from prescribed initial data (cf. \cite{ycb} and references therein). In particular, for the so-called cosmological case, at least  since the  investigations of Lifshitz and Khalatnikov in relativistic cosmology \cite{lk63}, perturbative approaches to the question of genericity of cosmological solutions constructed asymptotically from given initial data,  have been advanced in various directions in general cosmological theory. This was the way to prove the existence of  vacuum, regular cosmological solutions in general relativity having the required number to qualify as general ones, and also the non existence of a general singular, radiation solution in the same context, cf. \cite{lk63,ll}, eventually leading to the realization that the initial state was of a more complex nature \cite{bkl1,bkl2,he}. These first results were based on an approximation technique that consisted of several steps such as writing down a suitable expansion of the metric, substituting it to the field equations and counting the number of free (or arbitrary) functions needed to make the whole scheme consistent.

The original perturbative approximation scheme mentioned above proved to be especially fertile, and in fact it is being exploited in various directions ever since. It was used by Starobinski to study solutions of the Einstein equations with a positive cosmological constant, especially with regard to the question of the asymptotic stability of de Sitter space used in inflationary scenarios \cite{star1,star2}. The formal series expansions were used  more recently in \cite{star3} to study the genericity question in relativistic cosmologies with a general equation of state $p=w\r$, where it was proved that in order to be able to construct  a general singular solution initially, certain restrictions on the fluid parameter $w$ are needed. The ultrastiff case was recently considered in a more rigorous way using Fuchsian techniques  in Ref. \cite{heinzle}. Formal series where also used in \cite{star4} to study the more involved problem of perturbing an FRW universe containing two suitable fluids, and in \cite{bct} where it was shown how to construct a sudden singularity with the  genericity properties of a general solution, a situation met previously only in the `no-hair' behaviour of inflation. Extensions  to certain higher order gravity theories have also been considered in a formal series context, especially in connection with the stability of de Sitter space under generic perturbations in these theories, cf. \cite{star5,star6}.

It is interesting that the original approximation scheme of formal perturbation series has been further refined and applied in various situations in cosmology, cf. \cite{stew,tom,der} and references therein. More recently, Rendall has put the original formal series expansion techniques used by Starobinski \cite{star1} for the stability of de Sitter space, in a more rigorous basis and was able to prove various interesting theorems concerning function counting and formal series solutions of the Einstein equations with a positive cosmological constant in an `initial value problem' spirit, cf. \cite{ren1} (see also \cite{ren2}).

In this paper, we are interested in the possible genericity of \emph{regular} solutions defined by  formal series expansions in the context of vacuum models in higher order gravity derived from the analytic lagrangian $R+\epsilon R^2$ theory. In particular, we prove that the higher order gravity field equations in vacuum admit a unique solution in the form of a regular formal power series expansion which contains the right number of free functions to qualify as a general solution of the system. This requires a careful function counting technique, and for this purpose it is necessary to develop a formulation of the theory as a system of evolution equations with constraints.

The plan of this paper is as follows. In the next Section, we write the higher order gravity field equations  as a system of first order (ADM-type) evolution equations and constraints, as in an initial value formulation of the theory, and show that this system has the Cauchy-Kovalevski property. This allows us to count the true degrees of freedom necessary for any analytic solution to be  a general one. In Section 3, we introduce a regular formal series representation of the metric and prove our main result about the existence and uniqueness of a regular generic perturbation in higher order gravity starting from given formal asymptotic data.  Finally in the last Section, we present our conclusions and further discussion of these results. In Appendix A, we give more details about certain crucial steps in the proof of the Cauchy-Kovalevski property of our main system, while in Appendix B we present the full expressions of certain curvature components in terms of asymptotic data. In the last Appendix C, we discuss the technical issue of why it is justified to stop without loss of generality the formal expansions at the fourth-order terms.

\section{Function counting}
In this Section, we are interested in counting the true degrees of freedom of the $R+\epsilon R^2$ theory in  vacuum. This requires a splitting formulation of the theory into evolution equations and constraints, which in turn relies on certain technical details that the resulting dynamical system has to satisfy for the whole scheme to be consistent. The main technical result is given at the end of this Section,  Theorem 2.1.  We then count the degrees of freedom of the theory. In the Appendix A, we give more details of the proof of the main theorem of this Section.

We consider a spacetime $(\mathcal{V},g)$ where $\mathcal{V}=\mathbb{R}\times\mathcal{M}$, with $\mathcal{M}$ being an orientable 3-manifold, the submanifolds
$\mathcal{M}_{t}=\{t\}\times \mathcal{M}$, $t\in\mathbb{R}$, are spacelike and $g$ is a Lorentzian metric, analytic and
with signature $(+,-,-,-)$\footnote{Our conventions are those of \cite{ll}.}. We take a \emph{Cauchy adapted frame} $e_i=(e_{0},e_{\a})$
with $e_{\a}$ tangent to the space slice $\mathcal{M}_{t}$ and
$e_{0}$ orthogonal to it. The \emph{dual coframe}
$\theta^{i}=(\theta^{0}=dt,\theta^{\a}=dx^{\a}+\beta^{\a}dt)$,
 where the tangent vector
$\beta^{\a}$ is the usual \emph{shift}, leads to the standard general form
of the metric $g$,
\begin{equation}\label{metric}
  ds^2=N^{2}dt^{2}-\bar{g}_{\a\b}(t)\left(
  dx^{\a}+\beta^{\a}dt\right)\left( dx^{\b}+\beta^{\b}dt\right) .
\end{equation}
Here $N$ is a positive function,  the \emph{lapse}, and we assume
that all metrics $\bar{g}_{\a\b}(t)$ are complete Riemannian metrics. We could continue using a bar to denote spatial
tensors\footnote{We generally follow closely the terminology of the fundamental treatise \cite{ycb}.}, in a Cauchy adapted frame we would have $\bar{g}_{\a\b}=-g_{\a\b}$ and
$\bar{g}^{\;\a\b}=-g^{\a\b}$. However,  we more frequently write $ \GG_{\GA\GB}=-g_{\GA\GB}$. Below we shall deal exclusively with the simplest gauge choice, $N=1, \b=0$, which means that $ g_{00}=1, g_{0\GA}=0$, and local coordinates are \emph{adapted} to the product structure on $\mathcal{M}_{t}$, $(x^i=(t,x^\a))$ (in this case, one sometimes speaks of a \emph{synchronous} system of local coordinates).

Our starting point is the vacuum field equations of the higher order gravity theory derived from an analytic lagrangian $f(R)$, that is the equations
\be
L_{ij}=f'(R)R_{ij} -\frac{1}{2}f(R)g_{ij} -\nabla_i\nabla_j f'(R) +g_{ij}\Box_g f'(R)=0.
\label{eq:FEs}
\ee
We shall be  concerned below with the quadratic theory $f(R)=R+\GE R^2$, for which,  the field equations (\ref{eq:FEs}) in a Cauchy adapted frame  split as follows:
\be
L_{00} =(1+ 2\GE R)R_{00} -\frac{1}{2}(1+\GE R)R +2\GE g^{\GA\GB} \nabla_\GA \nabla_\GB R =0,
\label{eq:Loo}
\ee
\be
L_{0\GA} =(1 +2\GE R)R_{0\GA} -2\GE \nabla_0 \nabla_\GA R =0,
\label{eq:Loa}
\ee
\be
L_{\GA\GB} = (1 +2\GE R)R_{\GA\GB} -\frac{1}{2}(1+\GE R)R g_{\GA\GB} -2\GE \nabla_\GA \nabla_\GB R +2\GE g_{\GA\GB} \Box_g R =0.
\label{eq:Lab}
\ee
The components of the Ricci tensor are
\be
R_{00} = -\frac{1}{2}\partial_t K  -\frac{1} {4}K_\GA^\GB K_\GB^\GA,
\label{eq:R00mixeddown}
\ee
\be
R_{0\GA}  = \frac{1}{2} (\nabla_\b K^\GB_\GA - \nabla_\a K),
\label{eq:R0amixeddown}
\ee
and also
\be
R_{\GA\GB}= P_{\GA\GB} +\frac{1} {2} \partial_t K_{\GA\GB} +\frac{1} {4}K K_{\GA\GB} -\frac{1}{2}K^\GG_\GA K_{\GB\GG},
\label{eq:Rabmixeddown}
\ee
where $K=\textrm{tr}K_{\a\b}.$
Here, the extrinsic curvature is defined by the \emph{first variational equation}
\be\label{eq:pg}
\partial_t \g_{\GA\GB} = K_{\GA\GB},
\ee
and $P_{\GA\GB}$ denotes the three-dimensional Ricci tensor associated with $\GG_{\GA\GB}$.
Further, we define the \emph{acceleration tensor}  $D_{\GA\GB}$ through the \emph{second variational equation}
\be
\partial_t K_{\GA\GB} = D_{\GA\GB},
\label{eq:pK}
\ee
and we also introduce the \emph{jerk tensor} (3rd order derivatives) $W_{\GA\GB}$ through the \emph{jerk} equation:
\be
\partial_t D_{\GA\GB} = W_{\GA\GB}.
\label{eq:pD}
\ee
The space tensors $D_{\GA\GB}$ and $W_{\GA\GB}$ are obviously symmetric and because of the fourth order nature of the higher order field equations, they play an important role in what follows. Apart from the equations (\ref{eq:pg}) (velocity equation), (\ref{eq:pK}) (acceleration equation), (\ref{eq:pD}) (jerk equation), any initial data set $(\mathcal{M}_{t},\g_{\a\b},K_{\a\b},D_{\a\b},W_{\a\b})$ must satisfy Eq. (\ref{eq:Lab}). This results, after some manipulation,  in the following evolution equation, called the \emph{snap} equation:
\begin{eqnarray}
&&\partial_t W =\frac{1}{6\GE}( \frac{1}{2} P +\frac{1}{8} K^2  -\frac{5}{8} K^{\GA\GB} K_{\GA\GB} +D ) + \nonumber \\
&&\frac{1}{6}[P^2 +\frac{1}{4}P K^2 -\frac{1}{4}P K^{\GA\GB} K_{\GA\GB} +\frac{1}{32} K^4 -\frac{1}{16}K^2 K^{\GA\GB} K_{\GA\GB}  -\nonumber \\
&&6K K^\GA_\GB K^\GB_\GG K^\GG_\GA -\frac{99}{32} (K^{\GA\GB} K_{\GA\GB})^2 +27 K^\GA_\GB K^\GB_\GG K^\GG_\GD K^\GD_\GA +9 K K^{\GA\GB} D_{\GA\GB}  -  \nonumber \\
&&57 K^\GA_\GB K^\GB_\GG D^\GG_\GA  +\frac{13}{2}D K^{\GA\GB} K_{\GA\GB} -\frac{7}{2}D^2 +15 D^{\GA\GB} D_{\GA\GB}  -3K W   +\nonumber \\
&&15K^{\GA\GB} W_{\GA\GB}  -6\partial_t (\partial_t P) -\nonumber \\
&&4\g^{\GA\GB} \nabla_\GA \nabla_\GB (-P -D +\frac{3}{4} K^{\GA\GB} K_{\GA\GB} -\frac{1}{4}K^2)]
\label{eq:pW}
\end{eqnarray}

In terms of the acceleration tensor  $D_{\GA\GB}$, the Ricci tensor splittings given above become
\be
R_{00} = -\frac{1}{2}D +\frac{1} {4}K_\GA^\GB K_\GB^\GA,
\label{eq:R00mixeddownnew}
\ee
\be
R_{0\GA}  = \frac{1}{2} (\nabla_\b K^\GB_\GA - \nabla_\a K),
\label{eq:R0amixeddown}
\ee
\be
R_{\GA\GB}= P_{\GA\GB} +\frac{1} {2} D_{\GA\GB} +\frac{1} {4}K K_{\GA\GB} -\frac{1}{2}K^\GG_\GA K_{\GB\GG},
\label{eq:Rabmixeddownnew}
\ee
while for the scalar curvature we obtain,
\be
R = -P -D -\frac{1}{4}K^2 +\frac{3}{4}K^\GA_\GB K^\GB_\GA,
\label{eq:scalarR}
\ee
where $P=\textrm{tr}P_{\a\b},D=\textrm{tr}D_{\a\b}$.
If we substitute these forms into the identities (\ref{eq:Loo}),(\ref{eq:Loa}),  we find the following equations, which, in obvious analogy with the situation in general relativity, we call \emph{constraints}:

\emph{Hamiltonian Constraint}
\begin{eqnarray}
\mathcal{C}_0&=&\frac{1}{2}P  +\frac{1}{8}K^2 -\frac{1}{8}K^{\GA\GB} K_{\GA\GB} + \nonumber \\
&&\GE[-\frac{1}{2}P^2  -\frac{1}{4}P K^2 +\frac{1}{4}P K^{\GA\GB} K_{\GA\GB}  -\frac{1}{32}K^4 +\frac{1}{16}K^2 K^{\GA\GB} K_{\GA\GB} + \nonumber\\
&&\frac{3}{32}(K^{\GA\GB} K_{\GA\GB})^2 -\frac{1}{2}D K^{\GA\GB} K_{\GA\GB} +\frac{1}{2}D^2     - \nonumber\\
&&2\g^{\GA\GB} \nabla_\GA \nabla_\GB(-P - \frac{1}{4}K^2 +\frac{3}{4}K^{\GG\GD} K_{\GG\GD}  -D)] =0,
\label{eq:hamiltonian}
\end{eqnarray}

\emph{Momentum Constraint}
\begin{eqnarray}
\mathcal{C}_\a&=&\frac{1}{2} (\nabla_\b K^\GB_\GA - \nabla_\a K) + \nonumber \\
&&\GE[(-P  -\frac{1}{4}K^2 +\frac{3}{4}K^\GD_\GG K^\GG_\GD -D) (\nabla_\b K^\GB_\GA - \nabla_\a K)  - \nonumber \\
&&\nabla_\GA (-2\partial_t P + K K^{\GG\GD} K_{\GG\GD} -3 K^\GB_\GG K^\GG_\GD K^\GD_\GB -K D +5K^{\GG\GD} D_{\GG\GD} -2W)]=0
\label{eq:momentum}
\end{eqnarray}
The constraints show that the \emph{initial data} $(\g_{\a\b},K_{\a\b},D_{\a\b},W_{\a\b})$ cannot be chosen arbitrarily and must satisfy  the equations (\ref{eq:hamiltonian}) and (\ref{eq:momentum}) on each slice $\mathcal{M}_{t}$. Further, subtracting 4 diffeomorphisms, we find that  there are $24-4-4=16$ arbitrary functions to be specified initially.

The four evolution equations (\ref{eq:pg}), (\ref{eq:pK}), (\ref{eq:pD}) and (\ref{eq:pW}) are the higher order gravity analogues of the ADM equations of general relativity, and together with the constraints  (\ref{eq:hamiltonian}) and (\ref{eq:momentum}) describe the \emph{time development} $(\mathcal{V},g)$ of any initial data set $(\mathcal{M}_{t},\g_{\a\b},K_{\a\b},D_{\a\b},W_{\a\b})$ in higher order gravity theories.
However, for the function counting argument of this Section ($24-4-4=16$ arbitrary functions) to be verified, we need to prove that these equations are well defined in the sense that there is a well defined Cauchy problem, at least for the analytic case.
To be concrete, one needs to prove that these equations are of the Cauchy-Kovalevski type, that is there are no time derivatives in the constraints and the derivatives of the unknowns are (through the evolution equations)  analytic functions of the coordinates, the unknowns, and their first and second \emph{space} derivatives. This is shown in the Appendix. From this result and the Cauchy-Kovalevski theorem (cf. \cite{ycb}, Appendix V), we are immediately led to the following result, local Cauchy problem (analytic case):
 \begin{theorem}
 For $N=1,\b=0$, if we prescribe analytic initial data $(\g_{\a\b},K_{\a\b},D_{\a\b},W_{\a\b})$ on some initial slice $\mathcal{M}_{0}$, then there exists a neighborhood of $\mathcal{M}_{0}$ in $\mathbb{R}\times\mathcal{M}$ such that the evolution equations  (\ref{eq:pg}), (\ref{eq:pK}), (\ref{eq:pD}) and (\ref{eq:pW}) have an analytic solution in this neighborhood consistent with these data. This analytic solution is the development of  the prescribed initial data on  $\mathcal{M}_{0}$ if and only if these initial data satisfy the constraints.
  \end{theorem}
  The last part of this theorem follows most easily from the conformal equivalence theorem of higher order gravity theories \cite{ba-co88}, that is working in the Einstein frame representation, and a theorem on symmetric hyperbolic systems, cf. \cite{ycb}, pp. 150-1, and also \cite{ay}. Indeed, in the Einstein frame representation, the theory is general relativity plus a self-interacting scalar field, and the system becomes one of the form
\be
R^0_0 = -\frac{1}{2}\partial_t K -\frac{1} {4}K_\GA^\GB K_\GB^\GA = 8\pi k(T^0_0 -\frac{1}{2}T),
\label{eq:RooT}
\ee
\be
R^0_\GA  = \frac{1}{2} (\nabla_\b K^\GB_\GA - \nabla_\a K) = 8\pi k T^0_\GA,
\label{eq:RoaT}
\ee
\be
R^\GB_\GA = -P^\GB_\GA - \frac{1}{2\sqrt{\GG}} \partial_t(\sqrt{\GG}K^\GB_\GA) = 8\pi k (T^\GB_\GA -\frac{1}{2}\GD^\GB_\GA T),
\label{eq:RbaT}
\ee
and the wave equation $\nabla^i \nabla_i \phi -V'(\phi)=0$  for $\phi$ having the particular scalar field potential given in  \cite{ba-co88}. This system is of the form given in Thm. 4.1 of \cite{ycb}, p. 150, from which the result follows.

We note in passing that this result explains another function counting that is usually associated with higher order gravity theories, cf. \cite{star1,star2}, namely, that using the trace equation for the scalar curvature $R$ we get two initial data functions, and these together with the four other arbitrary functions coming from general relativity, result in the theory having finally 6  functions for the general solution in the analytic case. Although it may not be sound completely  right to count as independent functions  the metric functions together with the initial data corresponding to the trace equation for the scalar curvature and its first derivative  which in turn depend on the  metric and its derivatives, we may understand this result as follows. From the conformal transformation theorem above, we now that our higher order gravity theory is equivalent to general relativity plus a scalar field which satisfies a wave equation. Therefore in the conformal frame we indeed end up with 6 arbitrary functions, four from the geometric part and another two from the wave equation of the scalar field. However, the field $\phi$ is exactly defined to be directly related to the scalar curvature $R$ through the conformal transformation, $\phi=\ln (1+2\epsilon R)$, and so we see that the function counting resulting in the number 6 may be interpreted as giving the number of arbitrary functions associated with the conformal picture of the theory in the Einstein frame. We have shown in this section that the number of arbitrary functions of this same theory in the original Jordan frame is 16.

\section{Genericity of  regularity}
In this Section we perform a perturbative analysis of our basic system of higher order equations, namely, the evolution equations (\ref{eq:pg}), (\ref{eq:pK}), (\ref{eq:pD}) and (\ref{eq:pW}) together with the constraints  (\ref{eq:hamiltonian}) and (\ref{eq:momentum}). In particular, we assume a regular formal series representation of the spatial metric of the form
\be
\GG_{\GA\GB}= \g^{(0)}_{\GA\GB} +\g^{(1)}_{\GA\GB}\;t + \g^{(2)}_{\GA\GB}\;t^2 + \g^{(3)}_{\GA\GB}\;t^3 + \g^{(4)}_{\GA\GB}\;t^4 + \cdots
\label{eq:3dimmetric}
\ee
where the $ \g^{(0)}_{\GA\GB} , \g^{(1)}_{\GA\GB} , \g^{(2)}_{\GA\GB} , \g^{(3)}_{\GA\GB} , \g^{(4)}_{\GA\GB},\cdots$ are functions of the space coordinates. Because of the order of the higher order gravity equations, we shall be interested only in the part of the formal series shown, that is up to order four, and because of that we shall often drop the dots at the end of the various expressions to simplify the overall appearance\footnote{Differentiation of such formal series with respect to either space or the time variables is defined term by term, whereas multiplication of two such expressions results when the various terms are multiplied and terms of same powers of $t$ are taken together.}. Thus before substitution to the evolution and constraint higher order equations, the expression  (\ref{eq:3dimmetric}) contains  $30$ degrees of freedom. Note that setting $\g^{(0)}_{\GA\GB}=\d_{\a\b} $ and $\g^{(n)}_{\GA\GB}=0,n>0,$ we have Minkowski space included here as an exact solution of the equations, and so our perturbation analysis covers also that case.

The problem we are faced with in this Section is what the initial number of thirty free functions becomes after the imposition of the higher order evolution and constraint equations, that is  how it finally compares with the 16 degrees of freedom that any general solution must possess as shown in the previous Section. Put it more precisely, given data $a_{\GA\GB} ,b_{\GA\GB} ,c_{\GA\GB} ,d_{\GA\GB} ,e_{\GA\GB}$, arbitrary, nontrivial analytic functions of the space coordinates, such that the coefficients $\g^{(\mu)}_{\GA\GB}, \mu=0,\cdots 4,$ are prescribed,
\be\label{free}
 \g^{(0)}_{\GA\GB}= a_{\GA\GB}, \,\,  \g^{(1)}_{\GA\GB}=b_{\GA\GB}, \,\,   \g^{(2)}_{\GA\GB}= c_{\GA\GB}, \, \,   \g^{(3)}_{\GA\GB}= d_{\GA\GB}, \,\,
     \g^{(4)}_{\GA\GB}= e_{\GA\GB},
\ee
how many of these data are truly independent when (\ref{eq:3dimmetric}) is taken to be a possible solution of the evolution equations  (\ref{eq:pg}), (\ref{eq:pK}), (\ref{eq:pD}) and (\ref{eq:pW}) together with the constraints  (\ref{eq:hamiltonian}) and (\ref{eq:momentum})?

To proceed, we shall need the formal expansion of the reciprocal tensor $\GG^{\GA\GB}$, where $\GG_{\GA\GB}\GG^{\GB\GG}=\GD_\GA^\GG$. Using this, the various coefficients $ \g^{(\mu)\,\GA\GB},\mu=0,\cdots,4,$ of $t$ in the expansion $\g^{\GA\GB}=\sum_{n=0}^{\infty}\g^{(n)\,\GA\GB}t^n$ are found to be:
\bq
 \GG^{\GA\GB}&=& a^{\GA\GB} - b^{\GA\GB}t + \left(b^{\GA\GG}\;b_\GG^\GB - c^{\GA\GB}\right)t^2 +
 \left(-d^{\GA\GB} + b^{\GA\GG}\;c_\GG^\GB - b^{\GA\GG}\;b_\GG^\GD\;b_\GD^\GB +
 c^{\GA\GG}\;b_\GG^\GB\right)t^3   \nonumber     \\
&+& \left(-e^{\GA\GB} + b^{\GA\GG}d_\GG^\GB - b^{\GA\GG}b_\GG^\GD\;c_\GD^\GB + c^{\GA\GG}\;c_\GG^\GB + d^{\GA\GG}\;b_\GG^\GB - b^{\GA\GG}\; c_\GG^\GD b_\GD^\GB + b^{\GA\GG}b_\GG^\GD b_\GD^\GE b_\GE^\GB \right. \nonumber \\
&-& \left. c^{\GA\GG}\;b_\GG^\GD b_\GD^\GB\right)t^4.
\label{eq:3diminvmetric}
\eq
 Note that $a_{\GA\GB}a^{\GB\GG}=\GD_\GA^\GG$ and the indices of $b_{\GA\GB},c_{\GA\GB},d_{\GA\GB},e_{\GA\GB}$ are raised by $a^{\GA\GB}$.
For any tensor $X$, using the formal expansion  ($\ref{eq:3dimmetric}$), we can recursively calculate the coefficients in the expansion
\be
X_{\GA\GB}= X^{(0)}_{\GA\GB} +X^{(1)}_{\GA\GB}\;t + X^{(2)}_{\GA\GB}\;t^2 + X^{(3)}_{\GA\GB}\;t^3 + X^{(4)}_{\GA\GB}\;t^4 + \cdots,
\label{x tensor}
\ee
in particular we can write down a general iterated formula for the $n$-th order term, $X^{(n)}_{\GA\GB}$. For instance, for the extrinsic curvature $K_{\a\b}$, we write
 \be
K_{\GA\GB}= K^{(0)}_{\GA\GB} +K^{(1)}_{\GA\GB}\;t + K^{(2)}_{\GA\GB}\;t^2 + K^{(3)}_{\GA\GB}\;t^3.
\label{K tensor}
\ee
In terms of the data $a,b,c,,d,e$, we have explicitly,
\be
K_{\GA\GB}=\partial_t\GG_{\GA\GB}=b_{\GA\GB} + 2c_{\GA\GB}t + 3d_{\GA\GB}t^2 + 4e_{\GA\GB}t^3,
\label{eq:Kab}
\ee
and for the mixed components we obtain,
\bq
K_\GB^\GA&=&\GG^{\GA\GG}K_{\GG\GB}=b_\GB^\GA + (2c_\GB^\GA - b^{\GA\GG}b_{\GG\GB})t + (3d_\GB^\GA - 2b^{\GA\GG}c_{\GG\GB} + b^{\GA\GD} b_\GD^\GG  b_{\GG\GB} - c^{\GA\GG}b_{\GG\GB})t^2   \nonumber   \\
&+& (4e_\GB^\GA - 3b^{\GA\GG}d_{\GG\GB} + 2b^{\GA\GD}b_\GD^\GG c_{\GG\GB} - 2c^{\GA\GG}c_{\GG\GB} - d^{\GA\GG}b_{\GG\GB} + b^{\GA\GD}c_\GD^\GG b_{\GG\GB} - b^{\GA\GD}b_\GD^\GE b_\GE^\GG b_{\GG\GB} \nonumber\\
&+&  c^{\GA\GD}b_\GD^\GG b_{\GG\GB})t^3.
\label{eq:Kmixed}
\eq
Further, setting $\GG=|\GG_{\GA\GB}|=-g$, the mean curvature,
\be
K=K_\GA^\GA=\GG^{\GA\GB} \partial_t\GG_{\GA\GB}=\partial_t  ln(\GG) ,
\ee
is given by the form,
\bq
K&=&b + (2c - b_\GB^\GA b_\GA^\GB)t + (3d - 3b_\GB^\GA c_\GA^\GB + b^{\GA\GB}b_\GB^\GG b_{\GG\GA})t^2
\nonumber\\
&+& (4e -4b_\GB^\GA d_\GA^\GB + 4b^{\GA\GB}c_\GB^\GG b_{\GG\GA} - b^{\GA\GB}b_\GB^\GG b_\GG^\GD b_{\GD\GA}
-2 c_\GB^\GA c_\GA^\GB)t^3.
\label{eq:trK}
\eq
For completeness, we also give the expressions of the coefficients $K^{(n),\,\a\b}$ of the fully contravariant symbols,
\bq
K^{\a\b}&=&b^{\a\b} -2(b^{\a\g}b^\b_\g -c^{\a\b})t -3(-d^{\a\b} +b^{\a\g}c^\b_\g -b^{\a\g}b^\d_\g b^\b_\d +c^{\a\g}b^\b_\g)t^2
\nonumber\\
&-& 4(-e^{\a\b} +b^{\a\g}d^\b_\g -b^{\a\g}b^\d_\g c^\b_\d +c^{\a\g}c^\b_\g +d^{\a\g}b^\b_\g -b^{\a\g}c^\d_\g b^\b_\d\nonumber\\ &+&b^{\a\g}b^\d_\g b^\ep_\d b^\b_\ep -c^{\a\g}b^\d_\g b^\b_\d)t^3.
\label{eq:K^ab}
\eq
Using these forms, we can find the various components of the acceleration and jerk tensors to the required order. We have that the perturbation of the acceleration tensor in terms of the prescribed data is given by the form:
\bq
D_{\a\b}=\pa_t K_{\a\b}=2c_{\a\b} +6d_{\a\b}t +12e_{\a\b}t^2.
\label{eq:Dab}
\eq
Further we find,
\bq
D^\a_\b=\g^{\a\g}D_{\g\b}=2c^\a_\b +2(3d^\a_\b -b^\a_\g c^\g_\b)t +2(6e^\a_\b -3b^\a_\g d^\g_\b +b^\a_\d b^\d_\g c^\g_\b -c^\a_\g c^\g_\b)t^2,
\label{eq:Dmixed}
\eq
and,
\bq
D=2c +2(3d -b^\a_\b c^\b_\a)t +2(6e -3b^\a_\b d^\b_\a +b^\g_\b b^\b_\a c^\a_\g -c^\a_\b c^\b_\a)t^2,
\label{eq:trD}
\eq
where the trace is given by,
\[ D=D^\a_\a=\g^{\a\b} \pa_t K_{\a\b} . \]
For the fully contravariant components, we find,
\bq
D^{\a\b}&=& 2c^{\a\b} +2(3d^{\a\b} -b^{\a\g} c^\b_\g -b^{\d\b} c^\a_\d)t \nonumber\\
&+& 2(6e^{\a\b} -3b^{\a\g} d^\b_\g -3b^{\d\b} d^\a_\d -2c^{\a\g} c^\b_\g +b^{\a\d} b^\g_\d c^\b_\g +b^{\d\b} b^\a_\g c^\g_\d +b^{\d\g} b^\b_\g c^\a_\d)t^2.
\label{eq:D^ab}
\eq
Lastly, the jerk perturbation series is found to be,
\bq
W_{\a\b}=\pa_t D_{\a\b}= 6d_{\a\b} +24e_{\a\b}t,
\label{eq:Wab}
\eq
so that,
\bq
W= 6d +6(4e -b^\a_\b d^\b_\a)t,
\label{eq:trW}
\eq
where,
\[ W=W^\a_\a=\g^{\a\b} \pa_t D_{\a\b}.  \]
The components of the Ricci curvature  are more complicated when expressed in terms of the asymptotic data $a,b,c,d,e$ and we give them in the Appendix B. In terms of those and the various expressions obtained above, the hamiltonian constraint (\ref{eq:hamiltonian}) becomes,
\bq
\mathcal{C}_0&=&\frac{1}{2}P^{(0)} -\frac{1}{8}b^\b_\a b^\a_\b +\frac{1} {8}b^2
+ \ep \{2R^{(0)}(R^0_0)^{(0)}-\frac{1}{2}(R^{(0)})^2-2a^{\a\b}[\partial_\a \partial_\b R^{(0)}-\frac{1}{2}b_{\a\b}R^{(1)}-(\G^\m_{\a\b})^{(0)} \partial_\m R^{(0)})]\} \nonumber\\
&+& \{\frac{1}{2}P^{(1)}-\frac{1}{4}b^\b_\a b^\a_\b b+\frac{1}{4}b^\b_\a b^\a_\g b^\g_\b -\frac{1}{2}b^\b_\a c^\a_\b +\frac{1}{2}bc +\ep [2R^{(0)}(R^0_0)^{(1)} +2R^{(1)}(R^0_0)^{(0)}-R^{(0)}(R)^{(1)} \nonumber\\
&-& 2a^{\a\b}[\partial_\a \partial_\b R^{(1)}-c_{\a\b}R^{(1)}-b_{\a\b}R^{(2)}-(\G^\m_{\a\b})^{(1)} \partial_\m R^{(0)}-(\G^\m_{\a\b})^{(0)} \partial_\m R^{(1)}] \nonumber \\
&+& 2b^{\a\b}[\partial_\a \partial_\b R^{(0)}-\frac{1}{2}b_{\a\b}R^{(1)}-(\G^\m_{\a\b})^{(0)} \partial_\m R^{(0)})]\}t =0
\label{eq:L00mixed}
\eq
From (\ref{eq:momentum}), we calculate the momentum constraint in the form,
\bq
\mathcal{C}_\a&=&\frac{1} {2}(\nabla_\b b^\GB_\GA - \nabla_\a b) +
\ep\left[R^{(0)}(\nabla_\b b^\GB_\GA - \nabla_\a b) -2\partial_\a R^{(1)}+b^\b_\a \partial_\b R^{(0)}\right] \nonumber\\
&+& \left\{[(\nabla_\b c^\GB_\GA - \nabla_\a c) -\frac{1} {2}\nabla_\b (b_\GG^\GB b_\GA^\GG) + \frac{1} {2}\nabla_\a (b_\GG^\GB b_\GB^\GG)]\right.  \nonumber \\
&+& \GE\left[2R^{(0)}\left((\nabla_\b c^\GB_\GA - \nabla_\a c) -\frac{1} {2}\nabla_\b (b_\GG^\GB b_\GA^\GG) + \frac{1} {2}\nabla_\a (b_\GG^\GB b_\GB^\GG)\right) + R^{(1)}(\nabla_\b b^\GB_\GA - \nabla_\a b)  \right.
\label{eq:L0amixedet}\nonumber \\
&-& \left.\left.4\partial_\a R^{(2)} + b^\b_\a \partial_\b R^{(1)}+(2c^\b_\a -b^\b_\g b^\g_\a) \partial_\b R^{(0)}\right] \right\}t=0.
\label{eq:L0amixed}
\eq
Finally, the snap equation (\ref{eq:pW}) gives:
\bq
0&=&\frac{1}{2}P^{(0)}+2c-\frac{5}{8}b^\GB_\GA b^\GA_\GB+\frac{1}{8}b^2\nonumber\\
&+&\GE\left\{2[-P^{(0)}+(-2c+\frac{3}{4}b^\GD_\GG b^\GG_\GD-\frac{1}{4}b^2)][-P^{(0)}+(-c+\frac{1}{2}b^\GB_\GA b^\GA_\GB-\frac{1}{4}b^2)]\right.\nonumber\\
&-&\left. \frac{3}{2}[-P^{(0)}+(-2c+\frac{3}{4}b^\GD_\GG b^\GG_\GD-\frac{1}{4}b^2)]^2 \right. \nonumber\\
&+& \left.6[-2P^{(2)}-3bd-2c^2-24e+21b^\GG_\GD d^\GD_\GG+10c^\GG_\GD c^\GD_\GG-19b^\GE_\GD b^\GD_\GG c^\GG_\GE+\frac{9}{2}b^\GD_\GG b^\GG_\GE b^\GE_\GZ b^\GZ_\GD+3b^\GG_\GD c^\GD_\GG b\right.\nonumber\\
&+&\left.2b^\GG_\GD b^\GD_\GG c-b^\GG_\GD b^\GD_\GE b^\GE_\GG b-\frac{1}{2}(b^\GG_\GD b^\GD_\GG)^2]+4a^{\GA\GG}[\partial_\g(\partial_\a R^{(0)})-\frac{1}{2}b_{\g\a}R^{(1)}-(\G^\m_{\g\a})^{(0)}\partial_\m R^{(0)}]\right\}
\label{eq:trL}
\eq
To simplify our further work, we find it convenient to use the following  notation:
\bq
\mathcal{C}_0   & = & (L^0_0)^{(0)} + t(L^0_0)^{(1)} + \cdots
\label{eq:C0}
\eq
\bq
\mathcal{C}_\a  & = & (L^0_\a)^{(0)} + t(L^0_\a)^{(1)} + \cdots
\label{eq:Ca}
\eq
\bq
L^\b_\a & = & (L^\b_\a)^{(0)} + \cdots,
\label{eq:LAB}
\eq
using the formal expansion (\ref{x tensor}). For instance, the $O(t)$ term in the momentum constraint (\ref{eq:L0amixed}) is denoted by,
\bq
(L^0_\a)^{(1)}&=&[(\nabla_\b c^\GB_\GA - \nabla_\a c) -\frac{1} {2}\nabla_\b (b_\GG^\GB b_\GA^\GG) + \frac{1} {2}\nabla_\a (b_\GG^\GB b_\GB^\GG)]  \nonumber \\
&+& \ep\left[2R^{(0)}\left((\nabla_\b c^\GB_\GA - \nabla_\a c) -\frac{1} {2}\nabla_\b (b_\GG^\GB b_\GA^\GG) + \frac{1} {2}\nabla_\a (b_\GG^\GB b_\GB^\GG)\right) + R^{(1)}(\nabla_\b b^\GB_\GA - \nabla_\a b)  \right.
\label{eq:L0amixedet}\nonumber \\
&-& \left.4\partial_\a R^{(2)} + b^\b_\a \partial_\b R^{(1)}+(2c^\b_\a -b^\b_\g b^\g_\a) \partial_\b R^{(0)}\right].
\label{rel:Loa1}
\eq
We are now ready to decide how much the imposition of the field equations  (\ref{eq:pg}), (\ref{eq:pK}), (\ref{eq:pD}) and (\ref{eq:pW}) together with the constraints  (\ref{eq:hamiltonian}) and (\ref{eq:momentum}), restricts the number of free functions in the data (\ref{free}) in the perturbation series (\ref{eq:3dimmetric}). Using the field equations and calculating the various relations that appear at each order, we are led to the following relations between the data $a,b,c,d,e$:
From $(L^0_0)^{(0)}$, we get one relation, namely,
\bq
(L^0_0)^{(0)}&=&\frac{1}{2}P^{(0)} -\frac{1}{8}b^\b_\a b^\a_\b +\frac{1}{8}b^2
+\ep \{2(-P^{(0)} -2c +\frac{3}{4}b^\b_\a b^\a_\b -\frac{1} {4}b^2)(-c + \frac{1}{4}b_\a^\b b_\b^\a) \nonumber \\ &-&\frac{1}{2}(-P^{(0)} -2c +\frac{3}{4}b^\b_\a b^\a_\b -\frac{1} {4}b^2)^2
+ b(-bc -6d +5b^\b_\a c^\a_\b -\frac{3}{2}b^\b_\a b^\a_\g b^\g_\b +\frac{1}{2}b^\b_\a b^\a_\b b -P^{(1)}) \nonumber \\
&+&a^{\a\b}[-2\partial_\a \big( \partial_\b(-P^{(0)} -2c +\frac{3}{4}b^\b_\a b^\a_\b -\frac{1} {4}b^2) \big) \nonumber \\
&+& a^{\m\ep}A_{\a\b\ep}\partial_\m(-P^{(0)} -2c +\frac{3}{4}b^\b_\a b^\a_\b -\frac{1} {4}b^2)]\}=0,
\label{eq:P0}
\eq
where $A_{\a\b\ep}=\partial_\b a_{\a\ep} +\partial_\a a_{\b\ep} -\partial_\ep a_{\a\b}$.
From $(L^0_\a)^{(0)}$, we obtain three more relations,
\bq
(L^0_\a)^{(0)}&=&\frac{1}{2}(\nabla_\b b^\b_\a-\nabla_\a b)
+\ep \big[(-P_0 -2c +\frac{3}{4}b^\b_\a b^\a_\b -\frac{1} {4}b^2)(\nabla_\b b^\GB_\GA-\nabla_\a b) \nonumber \\
&-&2\partial_\a(-bc -6d +5b^\b_\a c^\a_\b -\frac{3}{2}b^\b_\a b^\a_\g b^\g_\b
+\frac{1}{2}b^\b_\a b^\a_\b b -P_1) + b^\b_\a \partial_\b(-P_0 -2c +\frac{3}{4}b^\b_\a b^\a_\b -\frac{1} {4}b^2) \big] \nonumber \\
&=&0,
\label{eq:gradb}
\eq
whereas from $(L^\b_\a)^{(0)}$, we get the following six relations,
\bq
(L^\b_\a)^{(0)}&=&-(P^\b_\a)^{(0)} -c^\b_\a +\frac{1} {2}b^\b_\g b^\g_\a -\frac{1} {4}b^\b_\a b -\frac{1}{2}\d^\b_\a(-P^{(0)}-2c+\frac{3}{4}b^\b_\a b^\a_\b -\frac{1}{4}b^2) \nonumber \\
&+&\ep \{2(-P^{(0)} -2c +\frac{3}{4}b^\b_\a b^\a_\b -\frac{1} {4}b^2)(-(P^\b_\a)^{(0)} -c^\b_\a +\frac{1} {2}b^\b_\g b^\g_\a -\frac{1} {4}b^\b_\a b) \nonumber \\
&-& b^\b_\a(-bc -6d +5b^\b_\a c^\a_\b -\frac{3}{2}b^\b_\a b^\a_\g b^\g_\b +\frac{1}{2}b^\b_\a b^\a_\b b -P^{(1)}) \nonumber \\
&+&a^{\b\g}[2\partial_\g \big(\partial_\a(-P^{(0)} -2c +\frac{3}{4}b^\b_\a b^\a_\b -\frac{1} {4}b^2)\big)  -a^{\m\ep}A_{\a\b\ep}\partial_\m(-P^{(0)} -2c +\frac{3}{4}b^\b_\a b^\a_\b -\frac{1} {4}b^2)] \nonumber \\
&+& \d^\b_\a[4\big(-\frac{3}{2}bd -c^2 -12e +\frac{21}{2}b^\b_\a d^\a_\b +5c^\b_\a c^\a_\b -\frac{19}{2}b^\b_\a b^\a_\g c^\g_\b  \nonumber \\
&+&\frac{9}{4}b^\b_\g b^\g_\a b^\a_\d b^\d_\b
+\frac{3} {2}b^\b_\a c^\a_\b b+  b^\b_\a b^\a_\b c -\frac{1} {2}b^\g_\b b^\b_\a b^\a_\g b -\frac{1} {4}(b^\b_\a b^\a_\b)^2 -P^{(2)}) \nonumber \\
&-&\frac{1}{2}(-P^{(0)} -2c +\frac{3}{4}b^\b_\a b^\a_\b -\frac{1} {4}b^2)^2
+ b(-bc -6d +5b^\b_\a c^\a_\b -\frac{3}{2}b^\b_\a b^\a_\g b^\g_\b +\frac{1}{2}b^\b_\a b^\a_\b b -P^{(1)}) \nonumber \\
&+&a^{\g\d}[\frac{1}{2}a^{\m\ep}A_{\g\d\ep}\partial_\m(-P^{(0)} -2c +\frac{3}{4}b^\b_\a b^\a_\b -\frac{1} {4}b^2) \nonumber \\
&-& 2\partial_\g \big(\partial_\d(-P^{(0)} -2c +\frac{3}{4}b^\b_\a b^\a_\b -\frac{1} {4}b^2) \big)] \big] \}=0.
\label{eq:cab}
\eq
Furthermore, we use the identity,
\bq
\nabla_i L^i_j=0,
\label{idLij}
\eq
for $j=0$ and $j=\a$. For $j=0$ we have,
\bq
\partial_t L^0_0-\g^{\a\b}\nabla_\b L^0_\a=0,
\label{idLio}
\eq
that is using (\ref{eq:3diminvmetric}),(\ref{eq:C0}) and (\ref{eq:Ca}), we find,
\bq
&&(L^0_0)^{(1)}-(\g^{\a\b})^{(0)}\nabla_\b (L^0_\a)^{(0)} \nonumber \\ &+&\sum^{n}_{k=1}\{[(k+1)(L^0_0)^{(k+1)}-\sum^{k}_{m=0}(\g^{\a\b})^{(m)}\nabla_\b (L^0_\a)^{(k-m)}]t^k\}+\cdots=0,
\label{idLioseries}
\eq
for any natural number $n\geq1$.
Taking into account the relation (\ref{eq:gradb}), for the zeroth-order term of the series (\ref{idLioseries}), we obtain,
\bq
(L^0_0)^{(1)}-(\g^{\a\b})^{(0)}\nabla_\b (L^0_\a)^{(0)}=0,
\label{idLio0}
\eq
so that the term $(L^0_0)^{(1)}$ vanishes identically.
For $j=\a$ we have,
\bq
\partial_t L^0_\a -\frac{1}{2}K^\b_\a L^0_\b +\nabla_\b L^\b_\a=0,
\label{idLia}
\eq
that is using (\ref{eq:Kmixed}), (\ref{eq:Ca}) and (\ref{eq:LAB}), we find,
\bq
&&(L^0_\a)^{(1)}-\frac{1}{2}(K^\b_\a)^{(0)}(L^0_\b)^{(0)}+\nabla_\b (L^\b_\a)^{(0)} \nonumber \\ &+&\sum^{n}_{k=1}\{[(k+1)(L^0_\a)^{(k+1)}-\frac{1}{2}\sum^{k}_{m=0}(K^\b_\a)^{(m)}(L^0_\b)^{(k-m)}
+\nabla_\b (L^\b_\a)^{(k)}]t^k\} +\cdots=0,
\label{idLiaseries}
\eq
for any natural number $n\geq1$. Taking into account the equations (\ref{eq:gradb}) and (\ref{eq:cab}), for the zeroth-order term of the series (\ref{idLia}), we obtain,
\bq
(L^0_\a)^{(1)}-\frac{1}{2}(K^\b_\a)^{(0)}(L^0_\b)^{(0)}+\nabla_\b (L^\b_\a)^{(0)}=0,
\label{idLia0}
\eq
that is, we find that the term $(L^0_\a)^{(1)}$ also vanishes identically.
Hence, in total we find that the imposition of the field equations leads to 10 relations between the
30 functions of the perturbation metric  (\ref{eq:3dimmetric}), that is we are left with 20 free functions. Taking into account the freedom we have in performing  4 diffeomorphism changes,  we finally conclude that there are in total 16 free functions in the solution  (\ref{eq:3dimmetric}). This means that  the regular solution (\ref{eq:3dimmetric}) corresponds to a general solution of the problem. Put it differently, regularity is a generic feature of the $R+\ep R^2$ theory in vacuum, assuming analyticity.

 We now ask: Out of the 30 different functions $a,b,c,d,e$, which sixteen of those are we to choose to use as our initial data? We have shown in this Section that the vacuum higher order gravity equations (\ref{eq:pg}), (\ref{eq:pK}), (\ref{eq:pD}) and (\ref{eq:pW}) together with the constraints  (\ref{eq:hamiltonian}) and (\ref{eq:momentum}), admit a regular formal series expansion of the form  (\ref{eq:3dimmetric})  as a general solution requiring 16 smooth initial data. If we prescribe the thirty data
 \be a_{\GA\GB}, \quad  b_{\GA\GB},\quad  c_{\GA\GB},\quad  d_{\GA\GB},\quad  e_{\GA\GB},
\ee
initially, we still have the freedom to fix 14 of them. We choose to leave the six components of the metric $a_{\GA\GB}$ free, and we choose the four symmetric space tensors $b_{\GA\GB},c_{\GA\GB},d_{\GA\GB}$ and $e_{\GA\GB}$ to be traceless with respect to $a_{\a\b}$. Then we proceed to count the number of free functions in several steps, starting from these $6+4\times 5=26$ functions. First,  (\ref{eq:P0}) fixes one of the components of $b_{\a\b}$ and Eq. (\ref{eq:gradb}) fixes 3 more components of $ b_{\a\b}$, thus leaving  $ b_{\a\b}$ with one component. Further, we use  the 6 relations in (\ref{eq:cab}) to completely fix the remaining 5 components of $c_{\a\b}$ and the last of $b_{\a\b}$. Summing up the free functions we found, we end up with
\be
\underbrace{6}_{\textrm{from}\ a_{\a\b}}+\underbrace{0}_{\textrm{from}\ b_{\a\b}}+\underbrace{0}_{\textrm{from}\ c_{\a\b}}+\underbrace{10}_{\textrm{from}{\ d_{\a\b}}\ \textrm{and}\ e_{\a\b}}=16
 \ee
  suitable free data as required for the solution to be a general one.
We thus arrive at the following result which summarizes what we have shown in this Section, and generalizes a theorem of Rendall \cite{ren1} for higher order gravity theories that derive from the lagrangian $R+\ep R^2$.
\begin{theorem}
Let $a_{\a\b}$ be a smooth Riemannian metric , $b_{\a\b},c_{\a\b},d_{\a\b}$ and $e_{\a\b}$  be symmetric smooth tensor fields which are traceless  with respect to the metric $a_{\a\b}$, i.e., they satisfy $b=c=d=e=0$.  Then there exists a formal power series expansion solution of the vacuum higher order gravity equations of the form (\ref{eq:3dimmetric}) such that:
\begin{enumerate}
\item It is unique
\item The coefficients $\g^{(n)\,\GA\GB}$ are all smooth
\item It holds that $\g^{(0)}_{\GA\GB}=a_{\a\b}$ and $\g^{(1)}_{\GA\GB}=b_{\a\b},$ $\g^{(2)}_{\GA\GB}=c_{\a\b}$, $\g^{(3)}_{\GA\GB}=d_{\a\b}$ and  $\g^{(4)}_{\GA\GB}=e_{\a\b}$.

\end{enumerate}
\end{theorem}
In the course of the proof of this result, uniqueness followed because all coefficients were found recursively, while smoothness follows because in no step of the proof did we found it necessary to lower the $\mathcal{C}^\infty$ assumption. We also note that $b_{\a\b}$ and $c_{\a\b}$ are necessarily transverse with respect to $a_{\a\b}$.

\section{Discussion}
In this paper, we have treated the problem of the existence of generic perturbations of the regular state in higher order gravity in vacuum that derives from the lagrangian $R+\ep R^2$. We have shown that there is a regular state of the theory in the form of a formal series expansion having the same number of free functions as those required for a general solution of the theory. This means that there exists an open set in the space of initial data of the theory that leads to a regular solution having the correct number of free functions to qualify as a general solution.

To achieve this, we have shown that there exists a  first order formulation of the theory with the Cauchy-Kovalevski property. This formulation of the quadratic theory $R+\ep R^2$  evolves an initial data set $(\mathcal{M},\g_{\a\b},K_{\a\b},D_{\a\b},W_{\a\b})$ through a set of the four evolution equations (\ref{eq:pg}), (\ref{eq:pK}), (\ref{eq:pD}) and (\ref{eq:pW}) and the two constraints  (\ref{eq:hamiltonian}) and (\ref{eq:momentum}), and builds the time development $(\mathcal{V},g)$. What we have proved is that if we start with an initial data set in which the metric has the asymptotic form (\ref{eq:3dimmetric}) and evolve, then we can build an asymptotic development  in the form of a formal series expansion which satisfies the evolution and constraint equations and has the same number of free functions as those of a general solution of the theory. In other words, we have shown that regularity is a generic feature of the $R+\ep R^2$ theory under the assumption of analyticity.

The results of this paper provide the necessary background for various further investigations that we carry out currently. For example, elsewhere  we plan to examine the generic perturbation problem of the well known radiation solution of the quadratic  theory and extend the results of this paper to any perfect fluid spacetime with equation of state $p=w\r$ in higher order gravity. It is also interesting to further compare these radiation perturbations  with the situation in vacuum in the context of higher order gravity. We also wish to extend our present results to the case of arbitrary lapse and shift, and to consider the problem of the present paper in the conformal frame. We know that in the case of a positive cosmological constant and a general perfect fluid source, similar results to those of this paper hold, cf. \cite{ren1}. However, the scalar field case, especially with the potential of the Einstein frame representation of an $f(R)$ theory, is to our knowledge an open problem.

\section*{Acknowledgements}
We are especially grateful to Yvonne Choquet-Bruhat and Antonios Tsokaros for discussions.

\section*{Appendix A: Proof of the Cauchy-Kovalevski property}
In this Appendix we present details of the proof that the four evolution equations (\ref{eq:pg}), (\ref{eq:pK}), (\ref{eq:pD}) and (\ref{eq:pW}) constitute a Cauchy-Kovalevski type system. Indeed, the only `dangerous' terms present in these equations are the three terms
\be\partial_t P,\quad\partial_t(\partial_t P), \quad\nabla_\alpha \nabla_\beta (-P -D +\frac{3}{4}K^{\gamma\delta} K_{\gamma\delta}-\frac{1}{4}K^2),\ee
present in the constraints and in the snap equation (\ref{eq:pW}).

The spatial connection coefficients are given by the obvious formula
\be
\Gamma^\mu_{\alpha\beta}=\frac{1}{2}\gamma^{\mu\epsilon}(\partial_\beta \gamma_{\alpha\epsilon} +\partial_\alpha \gamma_{\beta\epsilon} -\partial_\epsilon \gamma_{\alpha\beta}).\label{gammas}
\ee
We then have
\be\partial_t \Gamma^\mu_{\alpha\beta}= -\frac{1}{2}K^{\mu\epsilon}(\partial_\beta \gamma_{\alpha\epsilon} +\partial_\alpha \gamma_{\beta\epsilon} -\partial_\epsilon \gamma_{\alpha\beta}) +\frac{1}{2}\gamma^{\mu\epsilon}(\partial_\beta K_{\alpha\epsilon} +\partial_\alpha K_{\beta\epsilon}- \partial_\epsilon K_{\alpha\beta}),\label{dtgamma}     \ee
and
\begin{eqnarray}
\partial_t^2 \Gamma^\mu_{\alpha\beta}&=& - K^{\mu\epsilon}(\partial_\beta K_{\alpha\epsilon} +\partial_\alpha K_{\beta\epsilon}- \partial_\epsilon K_{\alpha\beta})\nonumber\\ &-&\frac{1}{2}(D^{\mu\epsilon}-2K^{\mu\eta} K^\epsilon_\eta)(\partial_\beta \gamma_{\alpha\epsilon} +\partial_\alpha \gamma_{\beta\epsilon} -\partial_\epsilon \gamma_{\alpha\beta}) \nonumber\\ &+&\frac{1}{2}\gamma^{\mu\epsilon}(\partial_\beta D_{\alpha\epsilon} +
\partial_\alpha D_{\beta\epsilon}- \partial_\epsilon D_{\alpha\beta}).\label{ddtgamma}
\end{eqnarray}
We also set
\be \G_{\e\a\b}=\frac{1}{2}(\pa_\b \g_{\a\e} +\pa_\a \g_{\b\e} -\pa_\e \g_{\a\b}), \ee
so that $\G^\m_{\a\b}=\g^{\m\e} \G_{\e\a\b}$, and further we set
\be Z_{\e\a\b}=\frac{1}{2}(\pa_\b K_{\a\e} +\pa_\a K_{\b\e} -\pa_\e K_{\a\b}),     \ee
and
\be H_{\e\a\b}=\frac{1}{2}(\pa_\b D_{\a\e} +\pa_\a D_{\b\e} -\pa_\e D_{\a\b}).       \ee
It then follows that the time derivatives of these `fully covariant symbols' satisfy
\bq
\pa_t \G_{\e\a\b}&=&Z_{\e\a\b},\\
\pa_t Z_{\e\a\b}&=&H_{\e\a\b},\\
\pa_t^2 \G_{\e\a\b}&=&H_{\e\a\b}.
\eq
Hence we conclude that the first time derivatives of the spatial
 connection coefficients depend only of $(\g_{\a\b},K_{\a\b})$ and their first \emph{spatial} derivatives, and the second  time derivatives of the spatial
 connection coefficients depend only of $(\g_{\a\b},K_{\a\b},D_{\a\b})$ and their first \emph{spatial} derivatives.

Now, the spatial Ricci tensor is given by
\be P_{\a\b}=\pa_\m \G^\m_{\a\b} -\pa_\b \G^\m_{\a\m} +\G^\m_{\a\b}\G^\ep_{\m\ep} -\G^\m_{\a\ep}\G^\ep_{\b\m},
\ee
and so its time derivative is calculated to be
\be \partial_t P_{\alpha\beta}= \partial_\mu (\partial_t \Gamma^\mu_{\alpha\beta}) -\partial_\beta (\partial_t \Gamma^\mu_{\alpha\mu}) +  \partial_t \Gamma^\mu_{\alpha\beta} \Gamma^\epsilon_{\mu\epsilon} +\Gamma^\mu_{\alpha\beta} \partial_t \Gamma^\epsilon_{\mu\epsilon} -\partial_t \Gamma^\mu_{\alpha\epsilon} \Gamma^\epsilon_{\beta\mu} -\Gamma^\mu_{\alpha\epsilon} \partial_t \Gamma^\epsilon_{\beta\mu}.  \ee
Then, for the spatial scalar curvature,
\be P=\g^{\a\b}P_{\a\b} ,
\ee
we find that
\be \partial_t P=-K^{\alpha\beta} P_{\alpha\beta}+\gamma^{\alpha\beta}\partial_t P_{\alpha\beta},\ee
and therefore the first of the dangerous terms finally reads:
\begin{eqnarray}
\pa_t P&=&K^{\a\b}(-\pa_\m \g^{\m\ep}\G_{\ep\a\b} - \g^{\m\ep}\pa_\m\G_{\ep\a\b} +\pa_\b \g^{\m\ep}\G_{\ep\a\m} + \g^{\m\ep}\pa_\b\G_{\ep\a\m} - \g^{\m\ep}\g^{\z\e}\G_{\ep\a\b}\G_{\e\m\z} +\g^{\m\e}\g^{\ep\z}\G_{\e\a\ep}\G_{\z\b\m}) \nonumber\\
&+& \g^{\a\b}[-\pa_\m K^{\m\ep}\G_{\ep\a\b} - K^{\m\ep}\pa_\m\G_{\ep\a\b} +\pa_\m \g^{\m\ep}Z_{\ep\a\b} + \g^{\m\ep}\pa_\m Z_{\ep\a\b} - \pa_\b K^{\m\ep}\G_{\ep\a\m} - K^{\m\ep}\pa_\b\G_{\ep\a\m} \nonumber\\
&+& \pa_\b \g^{\m\ep}Z_{\ep\a\m} + \g^{\m\ep}\pa_\b Z_{\ep\a\m} \nonumber\\
&+& \g^{\z\e}\G_{\e\m\z}(-K^{\m\ep}\G_{\ep\a\b} +\g^{\m\ep}Z_{\ep\a\b})
+\g^{\m\z}\G_{\z\a\b}(-K^{\ep\e}\G_{\e\m\ep} +\g^{\ep\e}Z_{\e\m\ep}) \nonumber\\
&-& \g^{\ep\z}\G_{\z\b\m}(-K^{\m\e}\G_{\e\a\ep} +\g^{\m\e}Z_{\e\a\ep})
-\g^{\m\z}\G_{\z\a\ep}(-K^{\ep\e}\G_{\e\b\m} +\g^{\ep\e}Z_{\e\b\m})].
\end{eqnarray}
This means that the first dangerous term is `purely spatial'. This result also implies that the third dangerous term is also purely spatial, for it is calculated to be of the form
\begin{eqnarray}
 \nabla_\alpha \nabla_\beta (-P -D +\frac{3}{4}K^{\gamma\delta} K_{\gamma\delta}-\frac{1}{4}K^2)&=& \partial_\alpha[\partial_\beta (-P -D +\frac{3}{4}K^{\gamma\delta} K_{\gamma\delta}-\frac{1}{4}K^2)] \nonumber\\ &-&\Gamma^\mu_{\alpha\beta}\partial_\mu (-P -D +\frac{3}{4}K^{\gamma\delta} K_{\gamma\delta}-\frac{1}{4}K^2)\nonumber \\
&-&\frac{1}{2}K_{\alpha\beta}(-\partial_t P-W +\frac{5}{2}K^{\alpha\beta}D_{\alpha\beta} \nonumber\\
&-&\frac{3}{2}K^{\alpha\gamma}K^\beta_\gamma K_{\alpha\beta} -\frac{1}{2}KD +\frac{1}{2}K K^{\alpha\beta} K_{\alpha\beta}),
\end{eqnarray}
that is trouble could only had arisen from the first dangerous term, which however as we showed above is purely spatial.

Lastly, since
\be
\partial_t(\partial_t P)=-(D^{\alpha\beta}-2K^{\alpha\gamma}K^\beta_\gamma)P_{\alpha\beta} - 2K^{\alpha\beta}\partial_t P_{\alpha\beta} +\gamma^{\alpha\beta}\partial_t(\partial_t P_{\alpha\beta}), \ee
and
\bq\partial_t^2 P_{\alpha\beta}&=&\partial_\mu (\partial_t^2 \Gamma^\mu_{\alpha\beta}) -\partial_\beta(\partial_t^2 \Gamma^\mu_{\alpha\mu}) +  \partial_t^2 \Gamma^\mu_{\alpha\beta} \Gamma^\epsilon_{\mu\epsilon} +\Gamma^\mu_{\alpha\beta} \partial_t^2 \Gamma^\epsilon_{\mu\epsilon} - \partial_t^2 \Gamma^\mu_{\alpha\epsilon} \Gamma^\epsilon_{\beta\mu} \nonumber\\
&-&  \partial_t^2 \Gamma^\epsilon_{\beta\mu}\Gamma^\mu_{\alpha\epsilon} +2 \partial_t \Gamma^\mu_{\alpha\beta}\partial_t\Gamma^\epsilon_{\mu\epsilon} -
2 \partial_t \Gamma^\mu_{\alpha\epsilon} \partial_t\Gamma^\epsilon_{\beta\mu},
\eq
we find that the second dangerous term depends on $(\g_{\a\b},K_{\a\b},D_{\a\b})$ and its first and second spatial derivatives,  namely, it has the form:
\begin{eqnarray}
\pa_t(\pa_t P)&=&(D^{\a\b}-2K^{\a\g}K^\b_\g)(-\pa_\m \g^{\m\ep}\G_{\ep\a\b} - \g^{\m\ep}\pa_\m\G_{\ep\a\b} +\pa_\b \g^{\m\ep}\G_{\ep\a\m} + \g^{\m\ep}\pa_\b\G_{\ep\a\m}\nonumber\\
&-& \g^{\m\ep}\g^{\z\e}\G_{\ep\a\b}\G_{\e\m\z} +\g^{\m\e}\g^{\ep\z}\G_{\e\a\ep}\G_{\z\b\m})\nonumber\\
&-&2K^{\a\b}[-\pa_\m K^{\m\ep}\G_{\ep\a\b} - K^{\m\ep}\pa_\m\G_{\ep\a\b} +\pa_\m \g^{\m\ep}Z_{\ep\a\b} + \g^{\m\ep}\pa_\m Z_{\ep\a\b} - \pa_\b K^{\m\ep}\G_{\ep\a\m} - K^{\m\ep}\pa_\b\G_{\ep\a\m} \nonumber\\
&+& \pa_\b \g^{\m\ep}Z_{\ep\a\m} + \g^{\m\ep}\pa_\b Z_{\ep\a\m} \nonumber\\
&+& \g^{\z\e}\G_{\e\m\z}(-K^{\m\ep}\G_{\ep\a\b} +\g^{\m\ep}Z_{\ep\a\b})
+\g^{\m\z}\G_{\z\a\b}(-K^{\ep\e}\G_{\e\m\ep} +\g^{\ep\e}Z_{\e\m\ep}) \nonumber\\
&-& \g^{\ep\z}\G_{\z\b\m}(-K^{\m\e}\G_{\e\a\ep} +\g^{\m\e}Z_{\e\a\ep})
-\g^{\m\z}\G_{\z\a\ep}(-K^{\ep\e}\G_{\e\b\m} +\g^{\ep\e}Z_{\e\b\m})] \nonumber\\
&+& \g^{\a\b}\{-\pa_\m(D^{\m\ep}-2K^{\m\g}K^\ep_\g)\G_{\ep\a\b} - (D^{\m\ep}-2K^{\m\g}K^\ep_\g)\pa_\m\G_{\ep\a\b}\nonumber\\
&-& 2\pa_\m K^{\m\ep}Z_{\ep\a\b}-2K^{\m\ep}\pa_\m Z_{\ep\a\b} + \pa_\m \g^{\m\ep}H_{\ep\a\b} + \g^{\m\ep}\pa_\m H_{\ep\a\b}\nonumber\\
&+& \pa_\b(D^{\m\ep}-2K^{\m\g}K^\ep_\g)\G_{\ep\a\m} + (D^{\m\ep}-2K^{\m\g}K^\ep_\g)\pa_\b\G_{\ep\a\m}\nonumber\\
&+& 2\pa_\b K^{\m\ep}Z_{\ep\a\m} + 2K^{\m\ep}\pa_\b Z_{\ep\a\m} - \pa_\b \g^{\m\ep}H_{\ep\a\m} - \g^{\m\ep}\pa_\b H_{\ep\a\m}\nonumber\\
&+& \g^{\z\e}[-(D^{\m\ep}-2K^{\m\g}K^\ep_\g)\G_{\ep\a\b}-2K^{\m\ep}Z_{\ep\a\b} + \g^{\m\ep}H_{\ep\a\b}]\G_{\e\m\z} \nonumber\\
&+& \g^{\m\z}[-(D^{\ep\e}-2K^{\ep\g}K^\e_\g)\G_{\e\m\ep}-2K^{\ep\e}Z_{\e\m\ep} + \g^{\ep\e}H_{\e\m\ep}]\G_{\z\a\b} \nonumber\\
&-& \g^{\ep\z}[-(D^{\m\e}-2K^{\m\g}K^\e_\g)\G_{\e\a\ep}-2K^{\m\e}Z_{\e\a\ep} + \g^{\m\e}H_{\e\a\ep}]\G_{\z\b\m}\nonumber \\
&-& \g^{\m\z}[-(D^{\ep\e}-2K^{\ep\g}K^\e_\g)\G_{\e\b\m}-2K^{\ep\e}Z_{\e\b\m} + \g^{\ep\e}H_{\e\b\m}]\G_{\z\a\ep}\nonumber \\
&+& 2(-K^{\m\ep}\G_{\ep\a\b}+\g^{\m\ep}Z_{\ep\a\b})(-K^{\z\e}\G_{\e\m\z}+\g^{\z\e}Z_{\e\m\z})\nonumber\\
&-& 2(-K^{\m\e}\G_{\e\a\ep}+\g^{\m\e}Z_{\e\a\ep})(-K^{\ep\z}\G_{\z\b\m}+\g^{\ep\z}Z_{\z\b\m})\}.
\end{eqnarray}
This completes the proof that the evolution equations are a Cauchy-Kovalevski system.

\section*{Apppendix B: Ricci curvarure in terms of the data $a,b,c,d,e$}
We give here the various components of the Ricci curvature and the space-space components of the field equation  (\ref{eq:Lab}) in terms of the data $a,b,c,d,e$. We have:
\bq
R^0_0 &=& -\frac{1}{2}D +\frac{1} {4}K_\GA^\GB K_\GB^\GA =
\left(-c + \frac{1}{4}b_\GA^\GB b_\GB^\GA\right) +
\left(-3d + 2b_\GA^\GB c_\GB^\GA - \frac{1}{2}b_\GA^\GB b_\GG^\GA b_\GB^\GG\right)t \nonumber  \\
&+& \left(-6e +\frac{9}{2}b_\GA^\GB d_\GB^\GA -\frac{7}{2}b_\GA^\GB b_\GG^\GA c_\GB^\GG + \frac{3} {4}b_\GG^\GB
b_\GA^\GG b_\GD^\GA b_\GB^\GD + 2c_\GA^\GB c_\GB^\GA\right)t^2 + \cdots,
\label{eq:R00mixed}   \eq
\bq
R^0_\GA &=& \frac{1}{2} (\nabla_\b K^\GB_\GA - \nabla_\a K)=
\frac{1}{2}\left(\nabla_\b b^\GB_\GA - \nabla_\a b\right) +
\left(\nabla_\b c^\GB_\GA - \nabla_\a c -\frac{1} {2}\nabla_\b (b_\GG^\GB b_\GA^\GG) +
\frac{1}{2}\nabla_\a (b_\GG^\GB b_\GB^\GG)\right)t   \nonumber   \\
&+& \left[\frac{3}{2}(\nabla_\b d^\GB_\GA - \nabla_\a d) - \nabla_\b (b^\GB_\GG c^\GG_\GA) +
\frac{3}{2}\nabla_\a (b^\GB_\GG c^\GG_\GB) - \frac{1}{2}\nabla_\b (c^\GB_\GG b^\GG_\GA)
+\frac{1}{2}\nabla_\b (b^\GB_\GD b^\GD_\GG b^\GG_\GA) -\frac{1}{2}\nabla_\a (b^\GD_\GB b^\GB_\GG b^\GG_\GD)\right]t^2
\nonumber \\
&+& \left[2(\nabla_\b e^\GB_\GA -\nabla_\a e) -\frac{3}{2}\nabla_\b (b^\GB_\GG d^\GG_\GA)
+2\nabla_\a (b^\GB_\GG d^\GG_\GB) -\frac{1}{2}\nabla_\b (d^\GB_\GG b^\GG_\GA) -
\nabla_\b (c^\GB_\GG c^\GG_\GA) +\nabla_\a (c^\GB_\GG c^\GG_\GB)    \right.\nonumber \\
&+& \left.\nabla_\b (b^\GB_\GD b^\GD_\GG c^\GG_\GA)
+\frac{1}{2}\nabla_\b (b^\GB_\GD c^\GD_\GG b^\GG_\GA)
+\frac{1}{2}\nabla_\b (c^\GB_\GD b^\GD_\GG b^\GG_\GA) -2\nabla_\a (b^\GD_\GB c^\GB_\GG b^\GG_\GD) \right.\nonumber\\
&-&\left.\frac{1}{2}\nabla_\b (b^\GB_\GD b^\GD_\GE b^\GE_\GG b^\GG_\GA) +
\frac{1}{2}\nabla_\a (b^\GD_\GB b^\GB_\GG b^\GG_\GE b^\GE_\GD)\right] t^3
+ \cdots,
\label{eq:R0amixed}
\eq
\bq
R^\GB_\GA &=&
-P^\GB_\GA -\frac{1} {4} KK^\GB_\GA
-\frac{1} {2}D^\b_\a +\frac{1}{2}K^\b_\g K^\g_\a= \left(-(P^\GB_\GA)^{(0)}
-c^\GB_\GA + \frac{1} {2}b^\GB_\GG b^\GG_\GA -\frac{1} {4}b^\GB_\GA b\right) \nonumber \\
&+& \left(-3d^\GB_\GA + 2b^\GB_\GG c^\GG_\GA - b^\GB_\GD b^\GD_\GG b^\GG_\GA + c^\GB_\GG b^\GG_\GA -
\frac{1}{2}b^\GB_\GA c +\frac{1} {4}b^\GD_\GG b^\GG_\GD b^\GB_\GA -
\frac{1}{2}c^\GB_\GA b +\frac{1} {4}b^\GB_\GG b^\GG_\GA b-(P^\b_\a)^{(1)}\right)t    \nonumber     \\
&+& \left(-6e^\GB_\GA +\frac{9} {2}b^\GB_\GG d^\GG_\GA -3b^\GB_\GD b^\GD_\GG c^\GG_\GA +3c^\GB_\GG c^\GG_\GA
 +\frac{3} {2}d^\GB_\GG b^\GG_\GA -\frac{3} {2}b^\GB_\GD c^\GD_\GG b^\GG_\GA +\frac{3} {2}b^\GB_\GD b^\GD_\GE
 b^\GE_\GG b^\GG_\GA -\frac{3} {2}c^\GB_\GD b^\GD_\GG b^\GG_\GA \right.   \nonumber  \\
&-& \frac{3} {4}d^\GB_\GA b +\frac{1} {2}b^\GB_\GG
 c^\GG_\GA b -\frac{1} {4}b^\GB_\GD b^\GD_\GG b^\GG_\GA b +\frac{1} {4}c^\GB_\GG b^\GG_\GA b -c^\GB_\GA c +
 \frac{1} {2}c^\GB_\GA b^\GD_\GG b^\GG_\GD +\frac{1} {2}b^\GB_\GG b^\GG_\GA c
 -\frac{1} {4}b^\GB_\GG b^\GG_\GA b^\GD_\GE b^\GE_\GD                \nonumber    \\
&-& \left.\frac{3} {4}b^\GB_\GA d +\frac{3} {4}b^\GB_\GA b^\GD_\GG c^\GG_\GD -
 \frac{1} {4}b^\GB_\GA b^\GD_\GE b^\GE_\GG b^\GG_\GD-(P^\b_\a)^{(2)}\right) t^2 +\cdots,
\label{eq:Rabmixed}
\eq
where $P_{\GA\GB}$ the Ricci tensor associated with $\GG_{\GA\GB}$.
The  scalar curvature becomes an expression of the form,
\be
R = R^{(0)} + R^{(1)} t + R^{(2)} t^2 + \cdots,
\label{eq:Rshort}
\ee
explicitly we have,
\bq
R&=&-P -\frac{1}{4}K^2 +\frac{3}{4}K^\b_\a K^\a_\b -D \nonumber\\
&=& \left(-P^{(0)} -2c +\frac{3}{4}b^\GB_\GA b^\GA_\GB -\frac{1} {4}b^2\right) +
\left(-bc -6d +5b^\GB_\GA c^\GA_\GB -\frac{3}{2}b^\GB_\GA b^\GA_\GG b^\GG_\GB +
\frac{1}{2}b^\GB_\GA b^\GA_\GB b-P^{(1)}\right)t   \nonumber  \\
&+& \left(-\frac{3}{2}bd -c^2 -12e +\frac{21}{2}b^\GB_\GA d^\GA_\GB +5c^\GB_\GA c^\GA_\GB -
\frac{19}{2}b^\GB_\GA b^\GA_\GG c^\GG_\GB +\frac{9}{4}b^\GB_\GG b^\GG_\GA b^\GA_\GD b^\GD_\GB +
\frac{3} {2}b^\GB_\GA c^\GA_\GB b\right.     \nonumber  \\
&+& \left. b^\GB_\GA b^\GA_\GB c -\frac{1} {2}b^\GG_\GB b^\GB_\GA b^\GA_\GG b -
    \frac{1} {4}(b^\GB_\GA b^\GA_\GB)^2-P^{(2)}\right) t^2 +\cdots.
\label{eq:trR}
\eq
Using these forms, the space-space components of the field equations in a Cauchy adapted frame, (\ref{eq:Lab}), become:
\bq
L^\GB_\GA &=& -(P^\GB_\GA)^{(0)} +(-c^\GB_\GA+\frac{1}{2}b^\GB_\GG b^\GG_\GA-\frac{1}{4}b^\GB_\GA b)-\frac{1}{2}[-P^{(0)}+(-2c+\frac{3}{4}b^\GD_\GG b^\GG_\GD-\frac{1}{4}b^2)]\GD^\GB_\GA\nonumber\\
&+&\GE\left\{2[-P^{(0)}+(-2c+\frac{3}{4}b^\GD_\GG b^\GG_\GD-\frac{1}{4}b^2)][ -(P^\GB_\GA)^{(0)} +(-c^\GB_\GA+\frac{1}{2}b^\GB_\GG b^\GG_\GA-\frac{1}{4}b^\GB_\GA b)]\right.\nonumber\\
&-&\frac{1}{2}[-P^{(0)}+(-2c+\frac{3}{4}b^\GD_\GG b^\GG_\GD-\frac{1}{4}b^2)]^2\GD^\GB_\GA+2a^{\GB\GG}[\partial_\g(\partial_\a R^{(0)})-\frac{1}{2}b_{\g\a}R^{(1)}-(\G^\m_{\g\a})^{(0)}\partial_\m R^{(0)}]\nonumber\\
&+&4R^{(2)}\d^\b_\a -2\d^\b_\a a^{\g\d}[\partial_\g(\partial_\d R^{(0)})-\frac{1}{2}b_{\g\d}R^{(1)}-(\G^\m_{\g\d})^{(0)}\partial_\m R^{(0)}] +\cdots=0.
\eq

\section*{Appendix C: The minimal-order terms in the formal series}
In this Appendix, we analyze the question of whether or not  the  result of counting  free functions  would be altered had we not used in the series in (\ref{eq:3dimmetric}) terms not up to the fourth order  but up to any finite  $n$-th power of $t$, where $n$ is a natural number with $n\geq 4$. Then, we would have 30 initial data from $a_{\a\b}, b_{\a\b}, c_{\a\b}, d_{\a\b},e_{\a\b}$ and also another  $6\times(n-4)$ initial data from the additional spatial matrices, giving $6n+6$ initial data in total. From Eqns. (\ref{eq:hamiltonian}) and (\ref{eq:momentum}), we note that $\mathcal{C}_0$ and $\mathcal{C}_\a$ are third-order differential equations with respect to the time $t$, and from (\ref{eq:pW}) that the snap equation is a fourth-order differential equation with respect to $t$.  Therefore Eqns. (\ref{eq:C0}), (\ref{eq:Ca}) and (\ref{eq:LAB}) become,
\bq
\mathcal{C}_0   & = & (L^0_0)^{(0)} + t(L^0_0)^{(1)} + \cdots + t^{n-3}(L^0_0)^{(n-3)} +\cdots
\label{eq:C0n}
\eq
\bq
\mathcal{C}_\a  & = & (L^0_\a)^{(0)} + t(L^0_\a)^{(1)} +\cdots + t^{n-3}(L^0_\a)^{(n-3)} +\cdots
\label{eq:Can}
\eq
\bq
L^\b_\a & = & (L^\b_\a)^{(0)}  + t(L^\b_\a)^{(1)}+ \cdots + t^{n-4}(L^\b_\a)^{(n-4)} +\cdots
\label{eq:LABn}
\eq
In order to have the same conclusion as in the case of $n=4$ (which is 16 arbitrary functions for the regular solution), we must show that for any $n$ we have $6n-14$ derived relations. We note that  from $(L^0_0)^{(0)},(L^0_\a)^{(0)} $ and from $(L^\b_\a)^{(i)}$, for $i=0,1,...,n-4$, there are  $6n-14$ such relations in total, and so we must prove  that the terms $(L^0_0)^{(j)}$ and $(L^0_\a)^{(j)}$ all vanish identically for any $j=1,2,...,n-3$.
Proceeding inductively, we have already shown that this stands true for $n=4$, due to the fact that the terms $(L^0_0)^{(1)}$ and $(L^0_\a)^{(1)}$ vanish identically by the identity (\ref{idLij}). Suppose that our statement stands when $n=k$, where k is a natural number, with $k\geq 4$, namely, that the terms $(L^0_0)^{(j)}$ and $(L^0_\a)^{(j)}$ vanish identically for any $j=1,2,...,k-3$. Then for $n=k+1$, for the $(k-3)$-order term we obtain from
(\ref{idLioseries}),
\bq
(k-2)(L^0_0)^{(k-2)} - \sum^{k-3}_{m=0}(\g^{\a\b})^{(m)}\nabla_\b (L^0_\a)^{(k-3-m)}=0,
\label{idL00k-2}
\eq
and from the $n=k$ step, due to the fact that $k-3-m\leq k-3$, we have that the term ${L^0_0}^{(k-2)}$ vanishes identically. Also, from (\ref{idLiaseries}) for the $(k-3)$-order term we obtain,
\bq
(k-2)(L^0_\a)^{(k-2)} - \frac{1}{2}\sum^{k-3}_{m=0}(K^\b_\a)^{(m)}(L^0_\b)^{(k-3-m)} + \nabla_\b (L^\b_\a)^{(k-3)}=0,
\label{idLoak-2}
\eq
which vanishes identically due to the fact that $k-3-m\leq k-3$. Therefore, without loss of generality, we can study the problem ofthe counting  of the arbitrary functions in the regular case using terms up to  the 4th  order in the metric expansion (\ref{eq:3dimmetric}).

\end{document}